\documentclass[fdp,a4paper,fleqn,finallayout]{w-art}

\usepackage{times,cite,w-thm}
\usepackage{graphicx}
\usepackage{hyperref}
\usepackage{epsfig}

\begin{document}
\title{Cooperativity in light scattering by cold atoms}


\author[T. Bienaim\'e]{Tom Bienaim\'e\inst{1}}
\address[\inst{1}]{Universit\'e de Nice Sophia-Antipolis, CNRS, Institut Non
Lin\'eaire de Nice, UMR 7335, F-06560 Valbonne, France}
\author[R. Bachelard]{Romain Bachelard\inst{2}}
\address[\inst{2}]{Instituto de F\'{i}sica de S\~{a}o Carlos, Universidade de S\~{a}o Paulo, 13560-970 S\~{a}o Carlos, SP, Brazil}
\author[N. Piovella]{Nicola Piovella\inst{3}}
\address[\inst{3}]{Dipartimento di Fisica, Universit\`a Degli Studi di Milano,
Via Celoria 16, I-20133 Milano, Italy} 
\author[R. Kaiser]{Robin Kaiser\inst{1}
\footnote{Corresponding author\quad
E-mail:~\textsf{robin.kaiser@inln.cnrs.fr},
            Phone: +33 (0)4 92 96 73 91,
            Fax: +33 (0)4 92 96 73 33}}

\date{\today}

\begin{abstract}
A cloud of cold $N$ two-level atoms driven by a resonant laser
beam shows cooperative effects both in the scattered radiation
field and in the radiation pressure force acting on the cloud
center-of-mass. The induced dipoles synchronize and the
scattered light presents superradiant and/or subradiant features.
We present a quantum description of the process in terms of a
master equation for the atomic density matrix in the scalar,
Born-Markov approximations, reduced to the single-excitation
limit. From a perturbative approach for weak incident field, we
derive from the master equation the effective Hamiltonian, valid
in the linear regime. We discuss the validity of the driven timed
Dicke ansatz and of a partial wave expansion for different optical thicknesses and we give analytical
expressions for the scattered intensity and the radiation pressure
force on the center of mass. We also derive an expression for
collective suppression of the atomic excitation and the scattered
light by these correlated dipoles.
\end{abstract}

\maketitle

\section{Introduction}

Cooperative scattering by large collections of resonant atoms has
been studied extensively for many years adopting either a
classical or a quantum description. Often a quantum formalism is
more convenient to describe the atom-light interaction. For
instance, superradiant emission can be obtained when many
independent atoms interact resonantly with photons, as studied in
the seminal work by Dicke \cite{Dicke54}. However, many features
can well be described with classical models for the atoms with a
polarizability and for the light field. Many features of Dicke
superradiance can thus be explained using classical theory. As
atoms appear to be excellent systems to study any possible
deviation from classical many body features, it is of general
interest to understand and to monitor cooperative effects, even at
a classical level, in a cloud of atoms. New intriguing effects can
arise, when fluctuations due to the coupling with the vacuum modes
can no longer be described by a classical field approach and the
atoms can also become entangled during the cooperative scattering.
Such cooperative scattering with atomic ensembles can appear in a
number of experimental situations in free space \cite{Javanainen}
or in cavities \cite{Raimond}.

Collective spontaneous emission at single excitation level \cite{Dicke54} has also been studied in resonant nuclear scattering of synchrotron radiation \cite{Trammell99, Rohlsberger10} and has recently
received growing interest with the study of single photon
superradiance from $N$ two-level atoms prepared by the absorption
of a single photon \cite{Scully06,Eberly06,Svidzinsky08,Svi10}. It
has been shown that the photon is spontaneously emitted in the
same direction as the incident field with a cooperative decay rate
proportional to $N$ and inversely proportional to the size of the
atomic cloud \cite{Svidzinsky08}. In a series of theoretical
\cite{Courteille10} and experimental \cite{Bienaime10,Bender10}
studies, the authors have recently addressed the question of the
quasi-resonant interaction of light with clouds of cold atoms,
bridging the gap from single atom behavior, with granularity
effects due to the discrete nature of the atomic distribution, to
a `mean-field' regime, where a continuous density distribution is
the relevant description and leads to cooperative effects. In
these works, the authors describe the collective atomic response under
continuous excitation of a low intensity light field using an
effective Hamiltonian model valid in the linear regime. From it, the average radiation pressure
force and the scattered intensity have been derived, observing the modification from the
single-atom values due to cooperativity \cite{Bienaime11}.

In this paper, we present an extended model based on a master
equation in the single-excitation approximation. Such model
provides a more complete description of the linear regime, useful
to investigate the distinctions between classical and quantum
features, even though a complete description of quantum
correlations in our system requires to take into account a larger
number of excitations \cite{Popescu,Bariani}. In particular, we
show that by a first-order perturbation expansion of the master
equation we recover the previously adopted linear model from
classical optics. We also derive an analytical expression showing
the cooperative suppression of atomic excitation and the related
fluorescence. This effect bears some common features with the
dipole blockade studied in Rydberg systems \cite{Tong, Singer,
Molmer}. In our case the long range $1/r$ dipole-dipole coupling
is at the origin of this important suppression even for dilute
clouds of cold atoms. We note that this suppression occurs in the
linear optics regime and is thus not related to a photon blockade.

\section{The master equation approach}

\subsection{The exact master equation for driven atoms}

\begin{figure}[t]
\centerline{{\includegraphics[height=4cm]{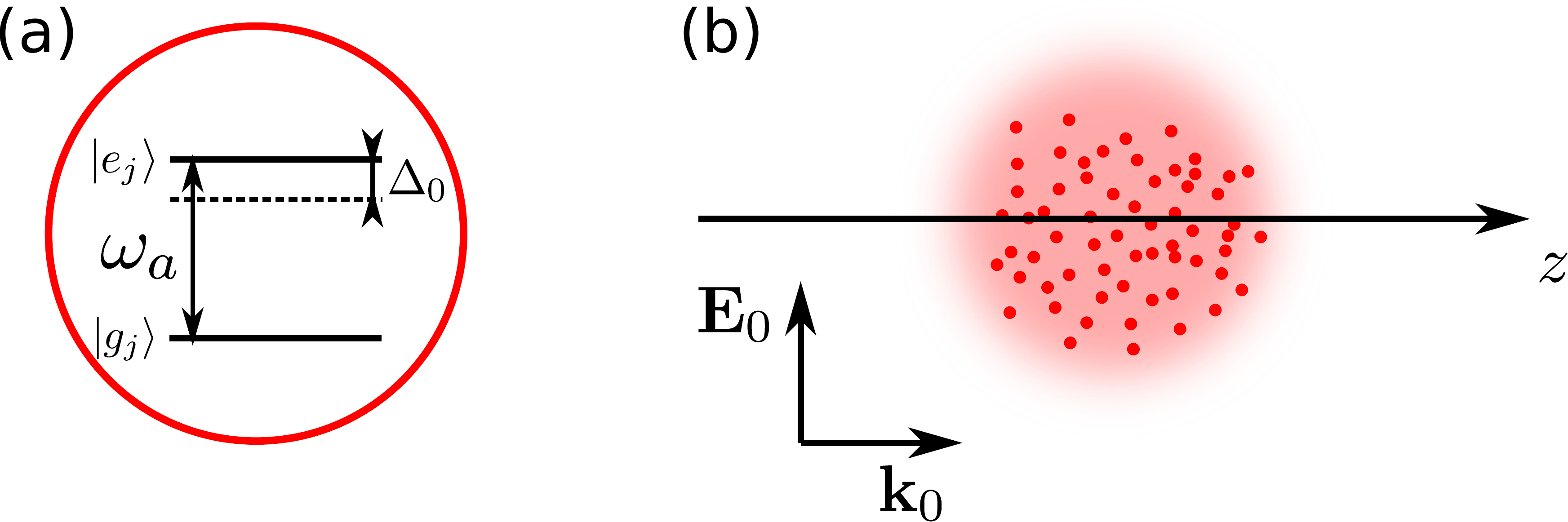}}}
\caption{(color online)  Experimental configuration: a cloud of
two-level atoms (a) is driven by an incident laser detuned by
$\Delta_0$ from the atomic resonance $\omega_a$, with wave vector
$\mathbf{k}_0$ (b).} \label{fig1}
\end{figure}

The dynamics of a system of atoms driven by an external laser beam
and undergoing cooperative re-emission into vacuum modes (see
Fig. \ref{fig1}) can be described by a master equation approach
\cite{Agarwal,Das}. Let us consider a system of $N$ two-level
atoms with transition frequency $\omega_a$, positions
$\mathbf{r}_j$ and excited decay time $\/\Gamma$. Each atom is
described by the spin half angular momentum algebra, with
$S^{i}_{-}=|g_i\rangle\langle e_i|$, $S^{i}_{+}=|e_i\rangle\langle
g_i|$, $S^{i}_{z}=|e_i\rangle\langle e_i|-|g_i\rangle\langle g_i|$
satisfying the commutation relations
$[S^{i}_{+},S^{j}_{-}]=\delta_{ij}S^{i}_{z}$ and
$[S^{i}_{\pm},S^{j}_{z}]=\mp 2\delta_{ij}S^{i}_{\pm}$.
The
interaction Hamiltonian is
\begin{eqnarray}\label{H}
    H&=&\frac{\hbar\Omega_0}{2}\sum_{j=1}^N
    \left(S_-^je^{i\Delta_0 t -i \mathbf{k}_0\cdot \mathbf{r}_j}+S_+^je^{-i\Delta_0 t +i \mathbf{k}_0\cdot
    \mathbf{r}_j}\right)\nonumber\\
   & & +\hbar\sum_{j=1}^N\sum_{\mathbf{k}}g_{\mathbf{k}}
    \left(S_-^je^{i\omega_0 t}+S_+^je^{-i\omega_0 t}\right)
    \left(a_{\mathbf{k}}e^{-i\omega_kt+i \mathbf{k}\cdot \mathbf{r}_j}+a_{\mathbf{k}}^\dagger
    e^{i\omega_kt-i \mathbf{k}\cdot \mathbf{r}_j}\right)
\end{eqnarray}
where $\Omega_0=dE_0/\hbar$ is the Rabi frequency of the classical
incident field with amplitude $E_0$ and wave vector
$\mathbf{k}_0$, $d$ is the dipole matrix element,
$\Delta_0=\omega_0-\omega_a$ is the pump-atom detuning,
$a_{\mathbf{k}}$ is the photon annihilation operator with wave
number $\mathbf{k}$ and frequency $\omega_k=ck$, $g_{\mathbf{k}} =
d[\omega_k/(2\hbar\epsilon_0 V_{ph})]^{1/2}$ and $V_{ph}$ the
photon volume.
We have assumed the rotating wave approximation (RWA) in the first
term accounting for the interaction with the external laser, but
not in the second term: there, the coupling between atoms and
vacuum field modes is described in the scalar light approximation,
where near field and polarization effects are neglected, since we
are considering dilute clouds, with $N(\lambda/R)^3<<1$ ($R$ the system size).

It has been shown in \cite{Agarwal} that the spontaneous emission
properties can be conveniently described by a reduced master
equation for the atomic system in the  Born-Markov approximation,
given by
\begin{eqnarray}\label{ME}
    \frac{d\rho}{dt}&=&-i\frac{\Omega_0}{2}\sum_{i}
    \left[e^{i\Delta_0 t-i\mathbf{k}_0\cdot \mathbf{r}_i}S^{i}_{-}+e^{-i\Delta_0 t+i\mathbf{k}_0\cdot
    \mathbf{r}_i}S^{i}_{+},\rho\right]\nonumber\\
   & & - i\sum_{i}\sum_{j\neq
    i}\Delta_{ij}[S^{i}_{+}S^{j}_{-},\rho]\nonumber\\
   & & +\frac{1}{2}\sum_{i}\sum_{j}\gamma_{ij}
    \left\{
    2S^{j}_{-}\rho S^{i}_{+}-S^{i}_{+}S^{j}_{-}\rho-\rho S^{i}_{+}S^{j}_{-}
    \right\}.
\end{eqnarray}
where
\begin{equation}\label{deltaij}
\Delta_{ij}=-\frac{\Gamma}{2}\frac{\cos(k_0|\mathbf{r}_i-\mathbf{r}_j|)}{k_0|\mathbf{r}_i-\mathbf{r}_j|},
\end{equation}
\begin{equation}\label{gammaij}
\gamma_{ij}=\Gamma\frac{\sin(k_0|\mathbf{r}_i-\mathbf{r}_j|)}{k_0|\mathbf{r}_i-\mathbf{r}_j|}.
\end{equation}
and $\Gamma = d^2k^3_0/(2\pi\hbar\epsilon_0)$.
 The first term in the master equation (\ref{ME}) describes the
interaction with the external laser.  The second term describes
the dipole-dipole interactions and arises from the virtual photon
exchange between pairs of atoms. It becomes especially important
at small interatomic distances, it is responsible for the
collective Lamb shift \cite{Friedberg,Scully09LS} and plays an
important role in the subradiant emission \cite{Svi10,Fano,Cummings}.
Finally, the third term of Eq. (\ref{ME}) describes the
cooperative emission. The Markov approximation ignores retardation
effects and takes the long time limit, i.e. $t\gg 1/\omega_0$ and
$t\gg \max\{|\mathbf{r}_{i}-\mathbf{r}_j|/c\}$. Hence, it requires
that the system size is not too large, such that one photon
travels through the atomic cloud faster than the characteristic
cooperative emission time.

The  RWA approximation adopted in the present model requires more
subtle arguments. As discussed in \cite{Agarwal}, when applied to
the second term of the Hamiltonian of Eq. (\ref{H}), the RWA consists
in ignoring anti-resonant terms like $a_{\mathbf{k}}^\dagger
S_+^j$ and $a_{\mathbf{k}} S_-^j$  which correspond to
simultaneous creation or annihilation of a photon and atomic
excitation (i.e. virtual transition). These terms are responsible
of the shift of the ground state which contributes to the terms
$\Delta_{ij}$ in Eq. (\ref{ME}). However, the master equation
(\ref{ME}) has been obtained in \cite{Agarwal} from the complete
Hamiltonian (\ref{H}) including the anti-resonant term, and making
the RWA only on the master equation, neglecting the rapidly
oscillating terms like $S_+^{j}\rho S_+^{i} e^{2i\omega_0 t}$.
Hence, Eq. (\ref{ME}) obtained by making the RWA on the master
equation rather than on the Hamiltonian does include the shift of
the ground state. These remarks make clear that RWA on the
Hamiltonian is not the same as RWA on the master equation and that
one should make RWA on the final equations of motion. Since the
counter-rotating terms as $S_+\rho S_+ e^{2i\omega_0 t}$ are not
important because $\Gamma\ll\omega_{0}$, Eq. (\ref{ME}) provides an
accurate description of the interaction, including both the
contributions of the real and virtual transitions to the frequency
shift.

\subsection{Single-excitation approximation}

The master equation (\ref{ME}) has been used to describe, in the
absence of the driving laser, the superradiant and subradiant
decays from excited atoms. For a system confined in a volume whose
size is much smaller than the radiation wavelength, and neglecting
the dipole-dipole interaction terms $\Delta_{ij}$, Dicke
\cite{Dicke54} introduced the angular momentum states
$|J,M\rangle=\textrm{Sym}\{|e\dots e;g\dots g\rangle\}$, with
$M=-J,\dots,J$. If the system is initially prepared in a symmetric
(superradiant) state (with e.g. $J=N/2$), without any coupling
between this state and the antisymmetric (subradiant) states. Note
however that the presence of dipole-dipole interactions may induce
coupling with subradiant states \cite{Fano}. In our case, the
presence of the driving laser provides the atom excitation which
subsequently scatters and/or decays cooperatively. If the driving
field is sufficiently weak or largely detuned from the atomic
resonance, we can assume that the atomic system is weakly excited,
with at most one atom out of $N$ excited. If the atoms are
organized in a symmetric state, the decay rate is $N\Gamma$.
Recently, it has been shown that an extended system of size $R \gg
\lambda$ ($\lambda=2\pi/k_0$ is the light wavelength) containing a
single excited atom among $N$ and prepared in the timed Dicke
state \cite{Scully06, Trammell99, Scully09,Svidzinsky08}
\begin{equation}\label{TD}
    |TD\rangle=\frac{1}{\sqrt{N}}\sum_{j=1}^N\exp(i\mathbf{k}_0\cdot
\mathbf{r}_j) \, |g_1,\dots,e_j,\dots,g_N\rangle,
\end{equation}
decays with a rate $\Gamma_N\propto N\Gamma/(k_0 R)^2$. Also, an external field $\Omega_0$
drives the system mainly in a steady state of the form
\cite{Courteille10,Bienaime10}
\begin{equation}\label{DTD}
    |\Psi\rangle_N\approx |g_1,\dots,g_N\rangle + \frac{\sqrt{N}\Omega_0e^{-i\Delta_0 t}}{2\Delta_0+i\Gamma(1+b_0/12)} |TD\rangle,
\end{equation}
where $b_0=(3\lambda^2/2\pi)\int dz \, \rho(0,0,z)\propto
N/(k_0R)^2$ is the resonant optical thickness along the
propagation direction $z$ of the driving laser and
$\rho(\mathbf{r})$ is the atomic density. However, a small
fraction of the initial ground state is still coupled to the
subradiant states, as discussed in \cite{Bienaime12}.

The previous works studying cooperative scattering by weakly
excited atoms were based on a effective Hamiltonian model and
brought us to a description equivalent to that of $N$ classical
linear dipoles driven by the external field
\cite{Courteille10,Bienaime11}. Here, we go beyond the linear
optics approximation using a master equation approach, still
restricted to a single excitation. This restriction would be
limited to the case of weak driving field, but may have a larger
validity, for instance for Rydberg's atoms where some kind of
blockade is provided \cite{Gaetan}.

By projecting Eq. (\ref{ME}) on the ground state
$|G\rangle\equiv|g_1,\dots,g_N\rangle$ and on the single-excitation
states $|i\rangle\equiv|g_1,\dots,e_i,\dots,g_N\rangle$, neglecting the
states containing more than one excitation and defining
$\rho_{i,G}=\langle i|\rho|G\rangle\exp(i\Delta_0 t)$,
$\rho_{G,G}=\langle G|\rho|G\rangle$ and $\rho_{i,j}=\langle
i|\rho|j\rangle$, we obtain
\begin{eqnarray}
  \frac{d\rho_{G,G}}{dt} &=& -i\frac{\Omega_0}{2}\sum_k
  \left(e^{-i\mathbf{k}_0\cdot \mathbf{r}_k}\rho_{k,G}-e^{i\mathbf{k}_0\cdot \mathbf{r}_k}\rho_{G,k}\right)
  +\sum_{k,l} \gamma_{kl}\rho_{l,k}\label{rho1},\\
  \frac{d\rho_{i,G}}{dt} &=&
  \left(i\Delta_0-\frac{\Gamma}{2}\right)\rho_{i,G}-i\frac{\Omega_0}{2}
  \left(e^{i\mathbf{k}_0\cdot \mathbf{r}_i}\rho_{G,G}-\sum_k e^{i\mathbf{k}_0\cdot \mathbf{r}_k}\rho_{i,k}\right)
  \nonumber \\
 & & -\sum_{k\neq i}\left(\frac{\gamma_{ik}}{2}+i\Delta_{ik}\right)\rho_{k,G}
   \label{rho2},\\
  \frac{d\rho_{i,j}}{dt} &=&  -i\frac{\Omega_0}{2}
  \left(e^{i\mathbf{k}_0\cdot \mathbf{r}_i}\rho_{G,j}-e^{-i\mathbf{k}_0\cdot \mathbf{r}_j}\rho_{i,G}\right)
  -i\sum_{k\neq i}\Delta_{ik}\rho_{k,j}+i\sum_{k\neq
  j}\Delta_{kj}\rho_{i,k} \nonumber  \\
 & & -\frac{1}{2}\sum_{k}\left(
  \gamma_{ik}\rho_{k,j}+\gamma_{kj}\rho_{i,k}\right).
   \label{rho3}
\end{eqnarray}
We note that these equations still conserve the probability:
$\rho_{G,G}+\sum_i\rho_{i,i}=1$.

\subsection{Perturbative solution}

We now show that the effective Hamiltonian approach can be
obtained by a perturbative expansion of $\rho$ in terms of the
Rabi frequency $\Omega_0$ of the external field. Assuming
$\rho=\rho^{(0)}+\rho^{(1)}+\dots$, if initially the system is in
the ground state, the zero-order term yields $\rho_{G,G}^{(0)}=1$
and $\rho_{i,G}^{(0)}=\rho_{i,j}^{(0)}=0$, whereas at the first
order, Eqs. (\ref{rho1})-(\ref{rho3}) reduce to:
\begin{eqnarray}
  \frac{d\rho_{G,G}^{(1)}}{dt}  &=& \sum_{k,l}\gamma_{kl}\rho_{l,k}^{(1)}, \label{rho1:1}\\
  \frac{d\rho_{i,G}^{(1)}}{dt} &=& \left(i\Delta_0-\frac{\Gamma}{2}\right)\rho_{i,G}^{(1)} -i\frac{\Omega_0}{2}e^{i\mathbf{k}_0\cdot \mathbf{r}_i}
  -\sum_{k\neq i}\left(\frac{\gamma_{ik}}{2}+i\Delta_{ik}\right)\rho_{k,g}^{(1)}, \label{rho2:1}\\
  \frac{d\rho_{i,j}^{(1)}}{dt} &=& -
  i\sum_{k\neq i}\Delta_{ik}\rho_{k,j}^{(1)}+i\sum_{k\neq
  j}\Delta_{kj}\rho_{i,k}^{(1)}  -\frac{1}{2}\sum_k
  \left(
  \gamma_{ik}\rho_{k,j}^{(1)}+\rho_{i,k}^{(1)}\gamma_{kj}
  \right)\label{rho3:1}.
\end{eqnarray}
Since $\rho_{i,j}(0)=0$, Eqs. (\ref{rho1:1}) and (\ref{rho3:1})
imply that at all times
$\rho_{i,j}^{(1)}(t)=\rho_{G,G}^{(1)}(t)=0$. The remaining Eq.
(\ref{rho2:1}) can be written defining $\beta_i=\rho_{i,G}^{(1)}$
and using Eqs. (\ref{deltaij}) and  (\ref{gammaij}), as
\begin{eqnarray}
  \frac{d\beta_i}{dt} &=& \left(i\Delta_0-\frac{\Gamma}{2}\right)\beta_i -i\frac{\Omega_0}{2}e^{i\mathbf{k}_0\cdot \mathbf{r}_i}
  -\frac{\Gamma}{2}\sum_{k\neq i}\frac{\exp(ik_0|\mathbf{r}_i-\mathbf{r}_k|)}{ik_0|\mathbf{r}_i-\mathbf{r}_k|}\beta_k.   \label{beta}
\end{eqnarray}
Furthermore, we observe that, since $\rho_{i,j}=\langle
i|\rho|G\rangle\langle G|\rho|j\rangle+\sum_k\langle
i|\rho|k\rangle\langle k|\rho|j\rangle$, the first non zero order
for the dipole correlations corresponds to the second order for the field: $\rho_{i,j}^{(2)}=\rho_{i,G}^{(1)}\rho_{G,j}^{(1)}=\beta_i\beta_j^*$.

\subsection{Effective Hamiltonian}

Eq. (\ref{beta}) can be expressed in the form of a Schr\"odinger
equation
\begin{equation}\label{eqS}
    i\hbar \frac{d|\Psi\rangle}{dt}=H_{\mathrm{eff}}|\Psi\rangle,
\end{equation}
where $|\Psi\rangle=\alpha|g\rangle+\sum_{i=1}^N\beta_i|i\rangle$
and the effective Hamiltonian is
\begin{equation}\label{Heff}
    H_{\mathrm{eff}}=\frac{\hbar\Omega_0}{2}\sum_i\left(e^{-i\mathbf{k}_0\cdot \mathbf{r}_i}S^{i}_{-}+e^{i\mathbf{k}_0\cdot
    \mathbf{r}_i}S^{i}_{+}\right)-\hbar\Delta_0\sum_i
    S^{i}_{+}S^{i}_{-}-\frac{\hbar\Gamma}{2}\sum_jV_{ij}S^{i}_{+}S^{j}_{-},
\end{equation}
where
\begin{equation}\label{Vij}
    V_{ij}=(1-\delta_{ij})\frac{\cos(k_0|\mathbf{r}_i-\mathbf{r}_j|)}{k_0|\mathbf{r}_i-\mathbf{r}_j|}+i
    \frac{\sin(k_0|\mathbf{r}_i-\mathbf{r}_j|)}{k_0|\mathbf{r}_i-\mathbf{r}_j|}.
\end{equation}
Assuming that at low saturation the system is weakly excited, so
that $\alpha\approx 1$, the projection of Eq. (\ref{eqS}) over the
excited states $|i\rangle$ leads to  Eq. (\ref{beta}). Is has been
shown that this equation describes also the temporal evolution of
$N$ harmonic oscillators driven by the scalar electric field
radiation $E_0$, as predicted by classical linear optics
\cite{Svi10,BienaimePhD}. It is important to notice that, even if the
collective atomic state $|\Psi\rangle$ is entangled, the knowledge
of only the probability amplitudes $\beta_i$ obtained from the
linear equations (\ref{beta}) is not by itself sufficient to detect
entanglement, due to its classical nature. Conversely,
entanglement could be observable in the solution of the master
equation, even restricted to a single-excitation, since it
contains correlations between atoms, and in particular between the
ground state and the excited states.

Finally, we notice that the exact master equation (\ref{ME}) can
be written in terms of the effective Hamiltonian (\ref{Heff}),
using Eqs. (\ref{eqS}), (\ref{Heff}) and
$\rho=|\Psi\rangle\langle\Psi|$, as
\begin{equation}\label{rhoHeff}
    \frac{d\rho}{dt}=\frac{1}{i\hbar}\left(
    H_{\mathrm{eff}}\rho-\rho
    H^{\dagger}_{\mathrm{\mathrm{eff}}}\right)+\sum_i\sum_j\gamma_{ij}S_-^j\rho
    S_+^i.
\end{equation}
The last term in Eq. (\ref{rhoHeff}) describes the refilling of the
ground state due to spontaneous decay of the excited states,
and is necessary to preserve the density operator
trace equal to unity.

\subsection{Timed Dicke state} \label{TDsection}

Let us show how the steady state Eq. (\ref{DTD}) provides an
approximated solution of the  single-excitation master equation.
First, we observe that the external field couples the ground state
$|G\rangle$ to the timed Dicke state $|TD\rangle$ defined in
Eq. (\ref{TD}). The matrix elements involving the TD state are
\begin{eqnarray}
    \rho_{TD,G}&=&\langle TD|\rho|G\rangle=\frac{1}{\sqrt{N}}\sum_{j}e^{-i\mathbf{k}_0\cdot
\mathbf{r}_j}\rho_{j,G}, \label{tdrho:1}\\
\rho_{TD,TD}&=&\langle
TD|\rho|TD\rangle=\frac{1}{N}\sum_{j}\sum_{m}
e^{-i\mathbf{k}_0\cdot
(\mathbf{r}_j-\mathbf{r}_m)}\rho_{j,m}.  \label{tdrho:2}
\end{eqnarray}
Their temporal evolutions are obtained from
Eqs. (\ref{rho1})-(\ref{rho3}) as
\begin{eqnarray}
  \frac{d\rho_{G,G}}{dt} &=& -i\frac{\sqrt{N}\Omega_0}{2}
  \left(\rho_{TD,G}-\rho_{G,TD}\right)
  +\sum_{k}\sum_{l} \gamma_{kl}\rho_{l,k}, \label{rho1:TD}\\
  \frac{d\rho_{TD,G}}{dt} &=&
  \left(i\Delta_0-\frac{\Gamma}{2}\right)\rho_{TD,G}-i\frac{\sqrt{N}\Omega_0}{2}
  \left(\rho_{G,G}-\rho_{TD,TD}\right) \nonumber \\
   & & -\frac{1}{\sqrt{N}}\sum_{i}e^{-i\mathbf{k}_0\cdot \mathbf{r}_i}\sum_{k\neq i}\left(\frac{\gamma_{ik}}{2}+i\Delta_{ik}\right)\rho_{k,G}, \label{rho:TD2}\\
  \frac{d\rho_{TD,TD}}{dt} &=&
  -i\frac{\sqrt{N}\Omega_0}{2}\left(\rho_{G,TD}-\rho_{TD,G}\right)\nonumber\\
  & & +\frac{i}{N}\sum_{i}\sum_{j}e^{-i\mathbf{k}_0\cdot(\mathbf{r}_i-\mathbf{r}_j)}
  \left(\sum_{k\neq i}\Delta_{kj}\rho_{i,k}-\sum_{k\neq
  i}\Delta_{ik}\rho_{k,j}\right)\nonumber\\
  & & - \frac{1}{2N}\sum_{i}\sum_{j}\sum_{k}e^{-i\mathbf{k}_0\cdot(\mathbf{r}_i-\mathbf{r}_j)}\left(
  \gamma_{ik}\rho_{k,j}+\gamma_{kj}\rho_{i,k}\right).
   \label{rho3:TD}
\end{eqnarray}
These equations show explicitly the coupling induced by the laser
between $|G\rangle$ and $|TD\rangle$. However, the dipole-dipole
interactions and the cooperative decay favor the coupling also to
all the other ``subradiant'' states $|s\rangle$, with
$s=1,\dots,N-1$, completing the single-excitation Hilbert
subspace. We make the assumption that all the matrix elements
among these states $|s\rangle$ can be neglected, supposing that
their occupation probability is small compared to the timed Dicke
state. In practice, we assume
\begin{equation}\label{rhoig}
    \rho_{i,G}=\langle i|TD\rangle\langle TD|\rho|G\rangle+\sum_s
    \langle i|s\rangle\langle s|\rho|G\rangle\approx \frac{1}{\sqrt{N}}
    e^{i\mathbf{k}_0\cdot\mathbf{r}_i}\rho_{TD,G},
\end{equation}
and
\begin{equation}\label{rhoij}
    \rho_{i,j}=\langle i|TD\rangle\langle TD|\rho|TD\rangle\langle TD|j\rangle+\dots
    \approx \frac{1}{N}
    e^{i\mathbf{k}_0\cdot(\mathbf{r}_i-\mathbf{r}_j)}\rho_{TD,TD}.
\end{equation}
By substituting these expressions in
Eq. (\ref{rho1:TD})-(\ref{rho3:TD}) we obtain
\begin{eqnarray}
  \frac{d\rho_{G,G}}{dt} &=& -i\frac{\sqrt{N}\Omega_0}{2}
  \left(\rho_{TD,G}-\rho_{G,TD}\right)
  +\Gamma_N\rho_{TD,TD}, \label{rho1:TD2}\\
  \frac{d\rho_{TD,G}}{dt} &=&
  \left[i\Delta_N-\frac{\Gamma_N}{2}\right]\rho_{TD,G}-i\frac{\sqrt{N}\Omega_0}{2}
  \left(\rho_{G,G}-\rho_{TD,TD}\right),
   \label{rho2:TD2}\\
  \frac{d\rho_{TD,TD}}{dt} &=&
  -i\frac{\sqrt{N}\Omega_0}{2}\left(\rho_{G,TD}-\rho_{TD,G}\right)-\Gamma_N\rho_{TD,TD},
   \label{rho3:TD2}
\end{eqnarray}
where $\Delta_N=\Delta_0-L_N$ and \cite{Bienaime11}
\begin{eqnarray}
  \Gamma_N &=& \frac{1}{N}\sum_{i}\sum_{j}\gamma_{ij}e^{-i\mathbf{k}_0\cdot(\mathbf{r}_i-\mathbf{r}_j)}
  =N\Gamma\langle |S_N(k_0,\theta,\phi)|^2\rangle_{\theta,\phi} \label{GammaN},\\
  L_N &=&-\frac{1}{N}\sum_{i}\sum_{j\neq i}\Delta_{ij}e^{-i\mathbf{k}_0\cdot(\mathbf{r}_i-\mathbf{r}_j)}
  =\frac{N\Gamma}{2\pi}\mathrm{P}\int_0^\infty d\kappa\frac{\kappa^3}{\kappa-1}\langle |S_N(k_0\kappa,\theta,\phi)|^2\rangle_{\theta,\phi},
\label{LN}
\end{eqnarray}
are the cooperative decay rate and the collective Lamb shift,
respectively. In  Eqs. (\ref{GammaN}) and (\ref{LN}) we introduced
the structure function
\begin{equation}\label{SN}
    S_N(\mathbf{k})=\frac{1}{N}\sum_{j=1}^N e^{i(\mathbf{k}-\mathbf{k}_0)\cdot \mathbf{r}_j}.
\end{equation}
and the average is taken over the total solid angle of emission of
a photon with wave vector $\mathbf{k}$ at an angle $\theta$ with
$\mathbf{k}_0$, where $|\mathbf{k}|=k_0$. Finally, the integral
over $\kappa$ in Eq. (\ref{LN}) is evaluated as a principal part.

\begin{figure}[t]
\begin{tabular}{cc}
\includegraphics[height=5cm]{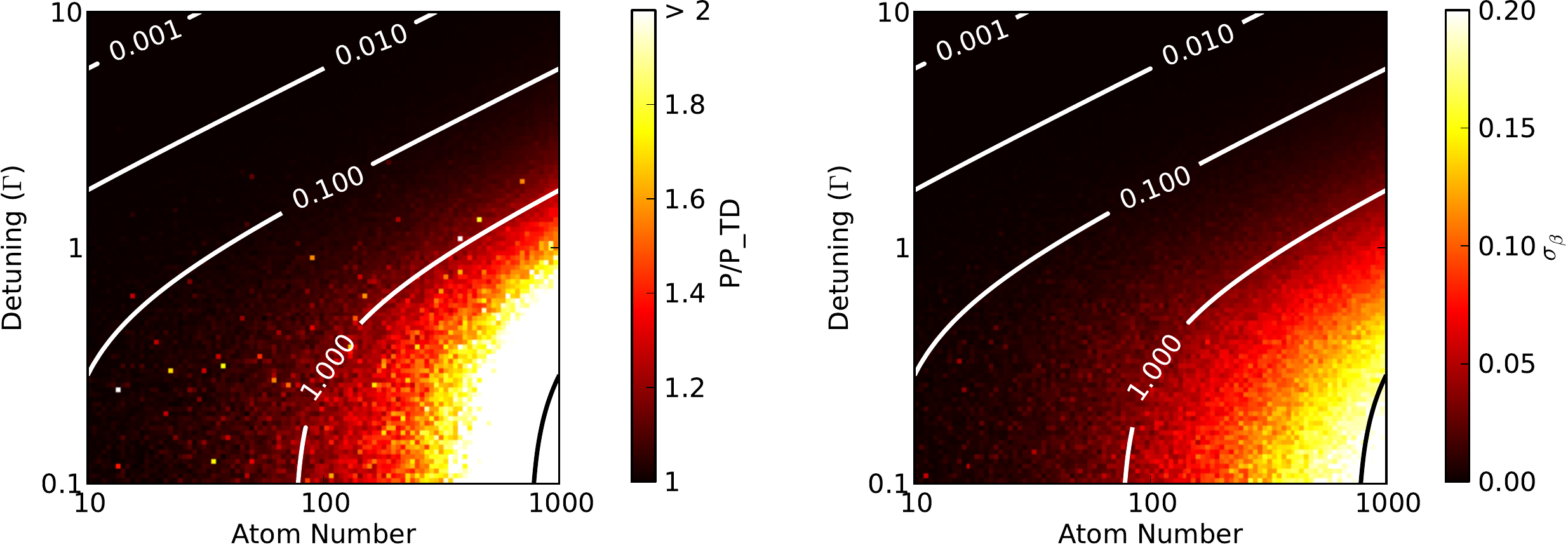}
\end{tabular}
\caption{(color online) Left: Ratio of the excited state population
computed numerically and in the timed Dicke approximation
$P/P_{TD}$. Right: Standard deviation
$\sigma_\beta$. The cloud is Gaussian with $k_0
\sigma_R = 15$. The contour plots show the iso-optical thickness
lines $b(\Delta_0)=b_0/(1+4 \Delta_0^2 / \Gamma^2)$ where $b_0 = 3
N / (k_0 \sigma_R)^2$. This color-coded plot helps visualize the
timed Dicke approximation validity region: $b(\Delta_0) < 1$. }
\label{fig2}
\end{figure}

Eqs. (\ref{rho1:TD2})-(\ref{rho3:TD2}) have the form of
Optical Bloch equations for a two-level system with collective
states $|G\rangle$ and $|TD\rangle$, interacting with a collective
Rabi frequency $\sqrt{N}\Omega_0$, detuning $\Delta_N$ and
linewidth $\Gamma_N$. The steady-state solution is given by,
\begin{eqnarray}
  \rho_{TD,G}^{s} &=& \frac{1}{1+s_c}\left(\frac{\sqrt{N}\Omega_0}{2\Delta_N+i\Gamma_N}\right),\\
  \rho_{TD,TD}^{s} &=& \frac{1}{2}\left(\frac{s_c}{1+s_c}\right),
\end{eqnarray}
where
\begin{equation}\label{sc}
    s_c=\frac{2N\Omega_0^2}{4\Delta_N^2+\Gamma_N^2},
\end{equation}
is the collective saturation parameter.
We recover the previous linear result of Ref.\cite{Courteille10}
in the limit $s_c\ll 1$, with a steady-state amplitude of the
excited state
\begin{equation}\label{betaTD}
    \beta_i\approx
    \left(\frac{\Omega_0}{2\Delta_N+i\Gamma_N}\right)e^{i\mathbf{k}_0\cdot\mathbf{r}_i}
    =\frac{\beta_{TD}}{\sqrt{N}}e^{i\mathbf{k}_0\cdot\mathbf{r}_i}.
\end{equation}
We have numerically verified that the timed Dicke ansatz of Eqs.
(\ref{rhoig}) and (\ref{rhoij}), or equivalently Eq.
(\ref{betaTD}) in the linear regime, is a good approximation of
the system when $b(\Delta_0) < 1$, where
$b(\Delta_0)=b_0/[1+(2\Delta_0/\Gamma)^2]$ is the optical
thickness. Figure \ref{fig2} shows a contour plot of the ratio
$P/P_{TD}$ ($P=\sum_j|\beta_j|^2$ and
$P_{TD}=|\beta_{TD}|^2$, where $\beta_j$ and $\beta_{TD}$ have been obtained from the numerical solution of Eqs.(\ref{beta}) and from (\ref{betaTD}) respectively) as a function of $N$ and
$\Delta_0/\Gamma$, for a Gaussian spherical distribution with
parameter $k_0\sigma_R=15$. For all values of $b(\Delta_0) < 1$,
the excited state population is well described by the timed Dicke
ansatz and yields a ratio $P/P_{TD}\sim 1$. This result is
confirmed by the analysis of the standard deviation $\sigma_\beta=\sqrt{\langle|\tilde\beta|^2\rangle-|\langle\tilde\beta\rangle|^2}/|\langle\tilde\beta\rangle|$
of $\tilde\beta=\beta
e^{-i\mathbf{k}_0.\mathbf{r}}$ (see right Fig.\ref{fig2} where the contour plot of $\sigma_\beta$ is also showed). $\sigma_\beta$ quantifies the
deviation of the excitation field from the `mean-field'
timed-Dicke state: this deviation becomes significant when $b(\Delta_0)$ is larger than unity. Fig.\ref{fig2} (right) suggests that for large optical thickness the homogeneous approximation assumed in the timed Dicke ansatz (see eq.(\ref{betaTD})) is no more valid and a better description is demanded.

\subsection{Beyond the timed Dicke state approximation} \label{BTDsection}

To account for the non-uniformity of the excitation within the cloud, the field  $\beta$ can be decomposed as a sum of waves that describe its spatial fluctuations. This approach has been considered
in Ref.\cite{Bachelard11}, where Eq.(\ref{beta}) has been solved
analytically for a continuous distribution with a Gaussian
spherical profile, neglecting the cosine part of the exponential
kernel and so the associated Collective Lamb shift $\Delta_{ij}$.
Although the use of a such truncated kernel is not allowed when
the decay of the excitation is observed \cite{Manassah12}, it may
still provide a reasonable approximation for small density and low
optical thickness $b(\Delta_0)$. The stationary excitation
amplitude obtained in ref.\cite{Bachelard11} using a partial wave
expansion is
\begin{equation}\label{betaBTD}
\beta(\mathbf{r})=\Omega_0\sum_{n=0}^\infty\frac{i^n(2n+1)}{2\Delta_0+i\Gamma(1+\lambda_n)}j_n(k_0r)P_n(\cos\theta),
\end{equation}
where $j_n$ are the spherical Bessel functions, $P_n$
the Legendre polynomials, $\theta$ the angle with respect to the
laser wave vector $\mathbf{k}_0$ and
$\lambda_n=N(\pi/2\sigma^2)^{1/2}I_{n+1/2}(\sigma^2)\exp(-\sigma^2/2)$,
where $\sigma=k_0\sigma_R$ and $\sigma_R$ is the rms width of the
Gaussian distribution. Eq. (\ref{betaBTD}) accounts for
non-homogeneity of the excitation probability density
$|\beta(\mathbf{r})|^2$. In this context, the timed Dicke expression (\ref{betaTD})
appears as a `mean-field' approximation, which can be recovered assuming
$\Gamma(1+\lambda_n)\sim \Gamma_N$ in Eq. (\ref{betaBTD}). Going further beyond, an
exact solution for the continuous-density limit of Eq.
(\ref{beta}) with the exponential kernel has been derived using the
Mie theory, although more mathematically demanding
\cite{Bachelard12}. In that case, the excitation of an atom was
calculated properly including both the incident and the
phase-shifted radiation field from the other atoms.

\section{Observables}
\subsection{Force on center of mass}

Cooperative effects can be investigated by a direct detection of
the scattered photons. However, this measurement can be in general
difficult, since for an extended atomic system the emission is
strongly forward directed and the detector can be saturated by the
incident laser. The scattered radiation field detected at distance
$\mathbf{r}$ and time $t$ is the sum of the single fields
scattered by the $N$ atoms of position $\mathbf{r}_j$,
\begin{equation}\label{ER}
    E(\mathbf{r},t)=\frac{d k_0^2}{2i\epsilon_0}\sum_{j=1}^N
    \frac{e^{-i\omega_0(t-|\mathbf{r}-\mathbf{r}_j|/c)}}{|\mathbf{r}-\mathbf{r}_j|}
    S_-^j(t).
\end{equation}
In the far field limit, $|\mathbf{r}-\mathbf{r}_j|\approx
r-(\mathbf{r}\cdot \mathbf{r}_j)/r$ and
\begin{equation}\label{E4}
    E(\mathbf{r},t)\approx\frac{d k_0^2}{2i\epsilon_0 r}e^{-i\omega_0(t-r/c)}\sum_{j=1}^N
    e^{-i\mathbf{k}\cdot \mathbf{r}_j}
    S_-^j(t),
\end{equation}
where $\mathbf{k}=k_0(\mathbf{r}/r)$ and $\omega_0=ck_0$, so that
the average intensity is
\begin{equation}\label{Isca}
    I(\mathbf{r},t)=\epsilon_0 c\langle E^\dagger(\mathbf{r},t)E(\mathbf{r},t)\rangle=
     \left(\frac{d^2\omega_0^4}{16\pi^2\epsilon_0 c^3 r^2}\right)\sum_{j}\sum_{m}
    e^{-i\mathbf{k}\cdot(\mathbf{r}_j-\mathbf{r}_m)}\rho_{j,m}(t).
\end{equation}
Alternatively, it is relatively easier to detect the cooperative
effects by considering the radiation pressure force exerted on the
atoms. If the atoms are sufficiently cold, it is experimentally
possible to measure the atomic motion after their exposition to
the incident laser beam. The radiation pressure force acting on
the $j$th-atom is $\hat{\mathbf{F}}_j=-\nabla_{\mathbf{r}_j}
H=\hat{\mathbf{F}}_{aj}+\hat{\mathbf{F}}_{ej}$  where
\cite{Courteille10,Bienaime11}
\begin{eqnarray}
    {\mathbf{F}}_{aj}&=& i\hbar \mathbf{k}_0\frac{\Omega_0}{2}
    \left\{e^{i\Delta_0t-i\mathbf{k}_0\cdot \mathbf{r}_j}S_-^j-e^{-i\Delta_0t+i\mathbf{k}_0\cdot \mathbf{r}_j}S_+^j\right\},\label{Force-abs}\\
    {\mathbf{F}}_{ej} &=&{\mathbf{F}}_{ej}^{(\mathrm{self})}
    -\frac{\hbar k_0\Gamma}{2}\sum_{m=1}^N
   \frac{\hat{\mathbf{r}}_{jm}}{(k_0 r_{jm})^2}
    \left\{S_+^j S_-^m(1-ik_0r_{jm})e^{ik_0r_{jm}}+h.c.\right\},
    \label{Force-emi}
\end{eqnarray}
where $r_{jm}=|\mathbf{r}_j-\mathbf{r}_m|=|\mathbf{r}_{jm}|$ and
$\hat{\mathbf{r}}_{jm}=\mathbf{r}_{jm}/r_{jm}$.
${\mathbf{F}}_{aj}$ and ${\mathbf{F}}_{ej}$ result from the recoil
received upon absorption of a photon from the pump and from the
emission of a photon into a direction $\mathbf{k}$, respectively.
The emission force $\mathbf{F}_{ej}$ acting on the $j$th-atom has
two contributions: a self-force
${\mathbf{F}}_{ej}^{(\mathrm{self})}=-\hbar\Gamma\sum_{|\mathbf{k}|=k_0}\mathbf{k}S_+^j
S_-^j$ due to its own photon emission, and a contribution
accounting for the coupling between the $j$th-atom and all the
other atoms. Note that the dipole-dipole interactions can occur
via a coupling to common vacuum modes of radiation. The
interference terms in the total scattered field can leave a
fingerprint on the forces acting on the atoms inside the cloud.
This force has a term decreasing as $1/r_{jm}$ and one decreasing
as $1/r_{jm}^2$. Their average values on the single-excitation atomic states are
\begin{eqnarray}
    \langle{\mathbf{F}}_{a}^j\rangle&=& i\hbar \mathbf{k}_0\frac{\Omega_0}{2}
    \left\{e^{-i\mathbf{k}_0\cdot \mathbf{r}_j}\rho_{j,G}-\textrm{c.c.}\right\}, \label{Force-abs2}\\
    \langle{\mathbf{F}}_{e}^j\rangle &=&
    -\frac{\hbar k_0\Gamma}{2}\sum_{m=1}^N
   \frac{\hat{\mathbf{r}}_{jm}}{(k_0 r_{jm})^2}
    \left\{\rho_{j,m}(1-ik_0r_{jm})e^{ik_0r_{jm}}+h.c.\right\}.
    \label{Force-emi2}
\end{eqnarray}
Notice that the self-force average to zero since the emission is
isotropic. The force on the center-of mass of the atomic cloud,
$\langle{\mathbf{F}}\rangle=(1/N)\sum_j
\langle{\mathbf{F}}_j\rangle$ is of particular interest. From
Eq. (\ref{Force-abs2}) and (\ref{Force-emi2}), its component along
the $z$ axis of incidence of the laser is
\begin{eqnarray}
    \langle{\mathbf{F}}_{z}\rangle&=& \frac{\hbar k_0}{N}\left\{
    \Omega_0\sum_j \textrm{Im}[\exp(i\mathbf{k}_0\cdot
    \mathbf{r}_j))\rho_{j,G}^*]-\Gamma\sum_{j,m}\hat z_{jm}j_1(k_0r_{jm})\textrm{Im}(\rho_{jm})\right\},
    \label{Force-CM}
\end{eqnarray}
where $j_1(z)=\sin(z)/z^2-\cos(z)/z$ is the first order spherical
Bessel function and $\hat{z}_{jm}=(z_j-z_m)/r_{jm}$. In the
timed-Dicke limit, using the approximations (\ref{rhoig}), (\ref{rhoij}) and neglecting saturation, Eq. (\ref{Force-CM}) becomes~\cite{Bienaime11}
\begin{eqnarray}
    \langle{\mathbf{F}}_{z}\rangle&=& \hbar k_0\Gamma
    \frac{\Omega_0^2}{4\Delta_N^2+\Gamma_N^2}N\left\langle (1-\cos\theta)|S_N(\mathbf{k})|^2\right\rangle_{\theta,\phi}.
    \label{Force-TD}
\end{eqnarray}

\begin{figure}[t]
\begin{tabular}{cc}
\includegraphics[height=3.0cm]{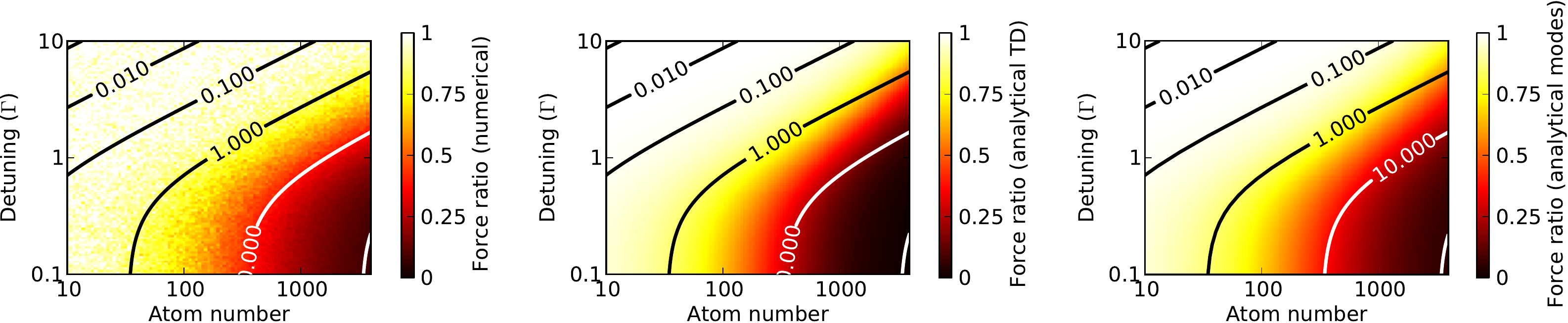}
\end{tabular}
\caption{(color online) Ratio of the cooperative to independent
radiation pressure force as a function of the atom number $N$ and
detuning $\Delta_0$. The cloud is Gaussian with $k_0 \sigma_R =
10$. The contour plots show the iso-optical thickness lines
$b(\Delta_0)=b_0/(1+4 \Delta_0^2 / \Gamma^2)$ where $b_0 = 3 N /
(k_0 \sigma_R)^2$. The left figure shows the numerical results
computed from the effective Hamiltonian Eq. (\ref{Heff}) and Eq.
(\ref{Force-CM}), the central figure stands for the analytical formula
Eq. (\ref{Force_Analytical}) and the right figure correspond to the modal expansion (\ref{ForcePRA}).} \label{fig3}
\end{figure}

The radiation pressure force can be influenced by different
effects. On one side, the finite extent of the atomic cloud can
produce strong forward oriented scattering. The balance between
the momentum of the incident and scattered photons and the atoms
indicate that for forward emission, the net recoil imprinted onto
the atoms is vanishing, resulting in a reduction of the radiation force. A
different contribution to the reduction of the radiation force can
be seen in the prefactor of Eq. (\ref{Force-TD}), which would
appear even in the case of isotropic scattering (i.e. when
$\langle \mathbf{F}_e\rangle=0$). The importance of this prefactor
can be understood from the cooperative coupling of several atoms
into the same vacuum mode. The number of available modes for large
spherical clouds can be estimated by $N_m \sim (k_0 R)^2$ (where
$R$ is the cloud's size), resulting in a number of atoms per mode
scaling as $N/N_m \sim N/(k_0 R)^2$. This scaling is conveniently
related to the on-resonant optical thickness of the atomic cloud
$b_0$. For a Gaussian density distribution with root mean square
size $\sigma_R$, $b_0=3N/(k_0 \sigma_R)^2$,
$N\langle|S_N|^2\rangle_{\theta,\phi}=1+b_0/12$ and
$N\langle(1-\cos\theta)|S_N|^2\rangle_{\theta,\phi}=b_0/24(k_0\sigma_R)^2$
\cite{Courteille10}.

It is convenient to compare the cooperative radiation pressure force to the force acting on a single independent atom
$F_{\text{ind}} = \hbar k_0 \Gamma \Omega_0^2 / (\Gamma^2 + \Delta_0^2)$. The ratio of the cooperative radiation pressure
force for a Gaussian cloud to the single-atom force in the timed Dicke limit can be written as
(neglecting the collective Lamb shift)
\begin{equation}
\frac{F_z}{F_{\text{ind}}} = \frac{4 \Delta_0^2 + \Gamma^2}{4 \Delta_0^2 + \left(1 + \frac{b_0}{12} \right)^2 \Gamma^2}
\left[1 + \frac{b_0}{24 (k_0 \sigma_R)^2 } \right]. \label{Force_Analytical}
\end{equation}
One can also use the partial wave expansion to account for the inhomogeneity of the field $\beta$, in which case the force ratio reads~\cite{Bachelard11}:
\begin{equation}
 \frac{F_z}{F_{\text{ind}}} =\frac{\Gamma^2+\Delta_0^2}{N}
\sum_{n=0}^\infty 
\left(\frac{(2n+1)\lambda_n(1+\lambda_n)}{4\Delta_0^2+\Gamma^2(1+\lambda_n)^2}
-\frac{(2n+2)\lambda_n\lambda_{n+1}[4\Delta_0^2+\Gamma^2(1+\lambda_n)(1+\lambda_{n+1})]}
    {[4\Delta_0^2+\Gamma^2(1+\lambda_n)^2][4\Delta_0^2+\Gamma^2(1+\lambda_{n+1})^2]}\right).\label{ForcePRA}
\end{equation}
Fig. \ref{fig3} shows the ratio of the cooperative to independent radiation pressure force as a function of the atom
number $N$ and detuning $\Delta_0$ for a Gaussian cloud with root mean square size $k_0 \sigma_R = 10$.
The left figure shows the numerical results computed from the effective Hamiltonian Eq. (\ref{Heff}) and Eq. (\ref{Force-CM}), the
central one describes the timed-Dicke formula Eq. (\ref{Force_Analytical}) and the right picture stands for the partial wave equation (\ref{ForcePRA}). While all methods predict a significant reduction of the force ratio for small detuning and large atom number, the partial wave approach yields better agreement with the numerical approach for $\Delta_0 < \Gamma$ and $N>1000$, despite the fact that the present partial wave approach has been limited to a sine kernel, not fully accounting for virtual-transition induced phase shifts~\cite{Manassah12}.

\subsection{Dicke subradiance}

\begin{figure}[t]
\centerline{{\includegraphics[height=12cm]{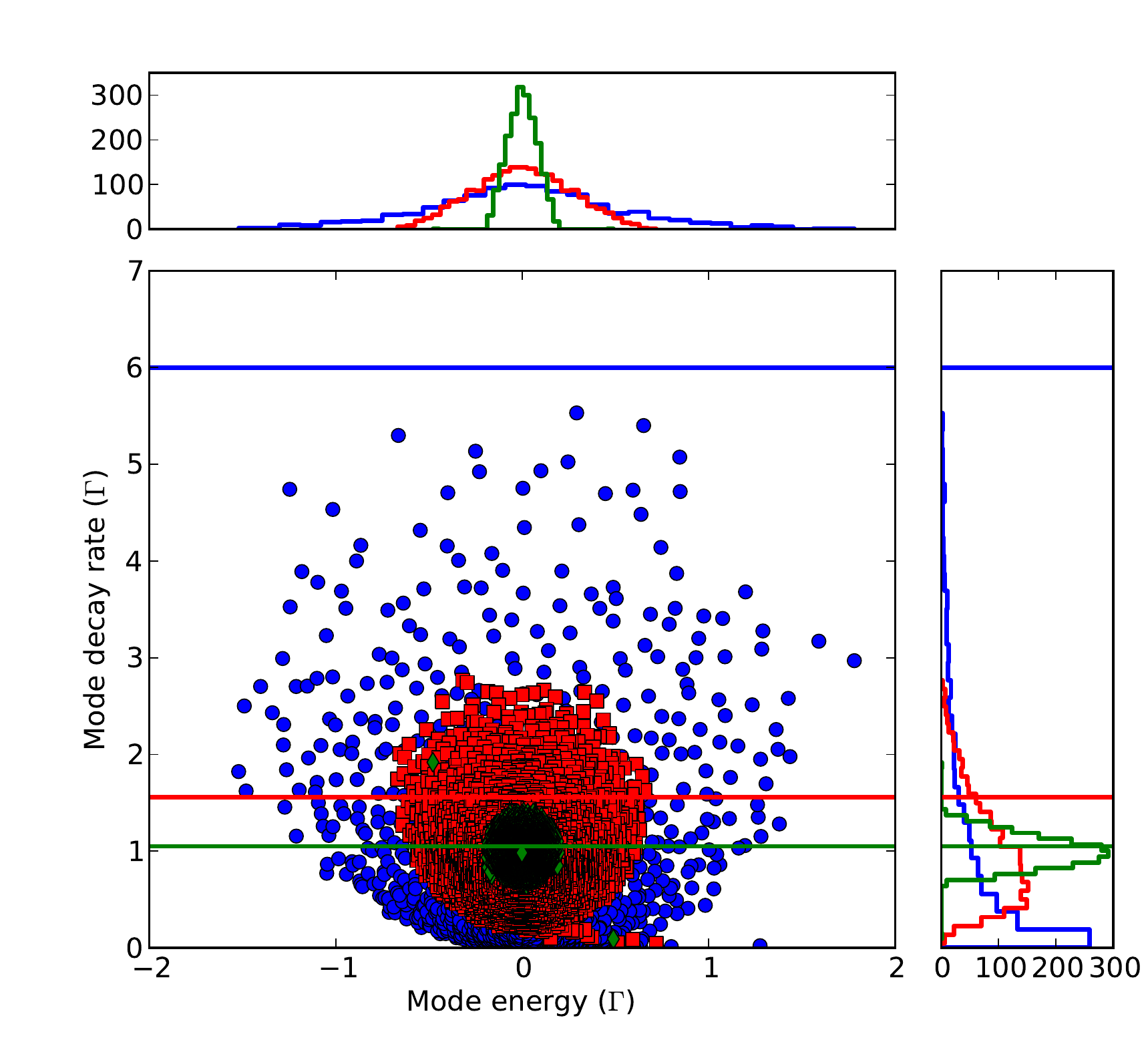}}}
\caption{(color online) Energies and decay rates of the modes of the system obtained by computing the eigenvalues of the effective Hamiltonian
$H_{\text{eff}}$ for Gaussian clouds with $N = 2000$ atoms and $k_0 \sigma_R = 10, 30, 100$ ($b_0 = 60, 6.7, 0.6$) given respectively by the blue,
red and green curves. The denser the system is, the larger the energy and decay rate distributions are.
The continuous straight lines show the timed Dicke state decay rates $\Gamma_N$ for the three different system sizes.
For dilute clouds (green line), the timed Dicke emission rate $\Gamma_N$ is centered on the distribution
$P(\Gamma)$ and $\Gamma_N \simeq \Gamma$. When the cloud optical density increases (blue line),
the timed Dicke decay rate tends to the tail of the distribution and $\Gamma_N \simeq \Gamma_{\text{max}}$.} \label{fig4}
\end{figure}

Dicke subradiance is the counterpart of superradiant emission and corresponds to the partial trapping of light due to destructive interferences.
In a subradiant state, the atomic dipoles are arranged such that the macroscopic polarization of the cloud is small reducing the emission rate
of the system. Subradiant emission has been previously observed for two ions \cite{Brewer96} and also for the emission of a cloud of $N$ atoms
in a free space into a single radiation mode \cite{Pillet85}. In a recent paper \cite{Bienaime12}, we showed that the system presented above is
ideal to observe for the first time long photon storage into metastable subradiant states for $N$ atoms in free space.

In section \ref{TDsection}, we saw that the laser pumps the system from the ground state $|G\rangle$ into the timed Dicke state $|TD\rangle$.
Then the dipole-dipole interaction terms couple the timed Dicke state to the different other states of the system. Some of these states
$|\psi_{\text{super}}\rangle$ have short lifetimes $\Gamma_{\text{super}} > \Gamma$ and are thus called \emph{superradiant}. Some other
states $|\psi_{\text{sub}}\rangle$ have long lifetimes $\Gamma_{\text{sub}} < \Gamma$ and are called \emph{subradiant}.
As the effective Hamiltonian is non-Hermitian, its eigenstates are not orthogonal and have common features with autoionizing states or Fano
resonances \cite{Fano}. Fig. \ref{fig4} shows the mode energies and decay rates for three different system sizes.
It shows that the denser the system is, the larger the energy and decay rate distributions are. Fig. \ref{fig4}
also shows the timed Dicke decay rate $\Gamma_N$ to compare it to the decay rate distribution of the system modes $P(\Gamma)$.
For dilute clouds, the timed Dicke emission rate $\Gamma_N$ is centered on the distribution $P(\Gamma)$ and $\Gamma_N \simeq \Gamma$
and when the cloud optical density increases, the timed Dicke decay rate tends to the tail of the distribution and
$\Gamma_N \simeq \Gamma_{\text{max}}$.

\begin{figure}[t]
\centerline{{\includegraphics[height=2.5cm]{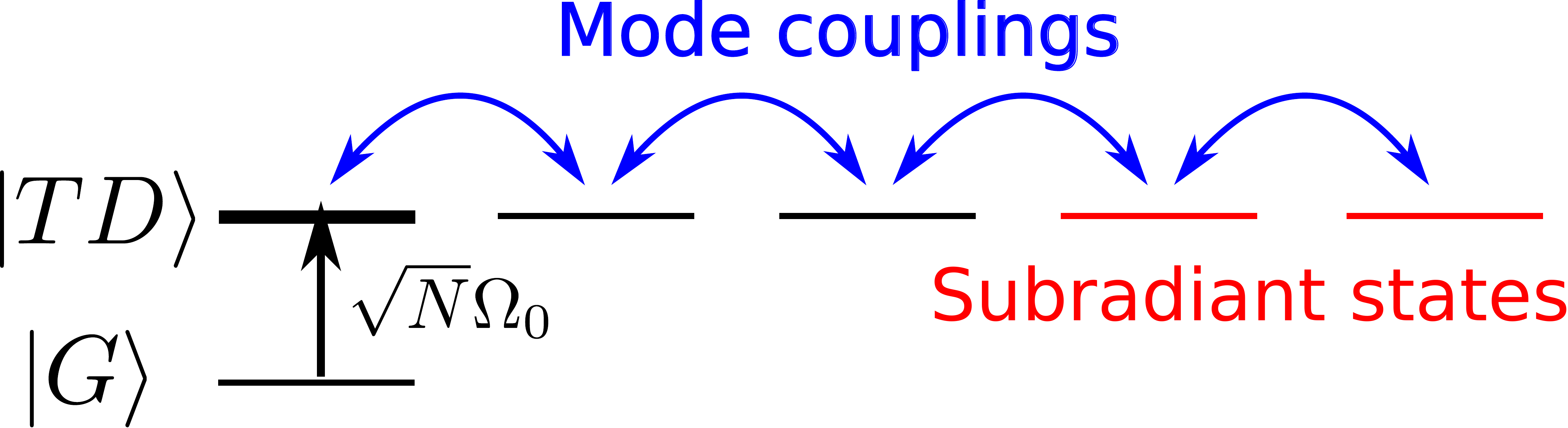}}}
\caption{(color online) Sketch of the subradiant emission couplings.} \label{fig5}
\end{figure}

\begin{figure}[t]
\centerline{{\includegraphics[height=5cm]{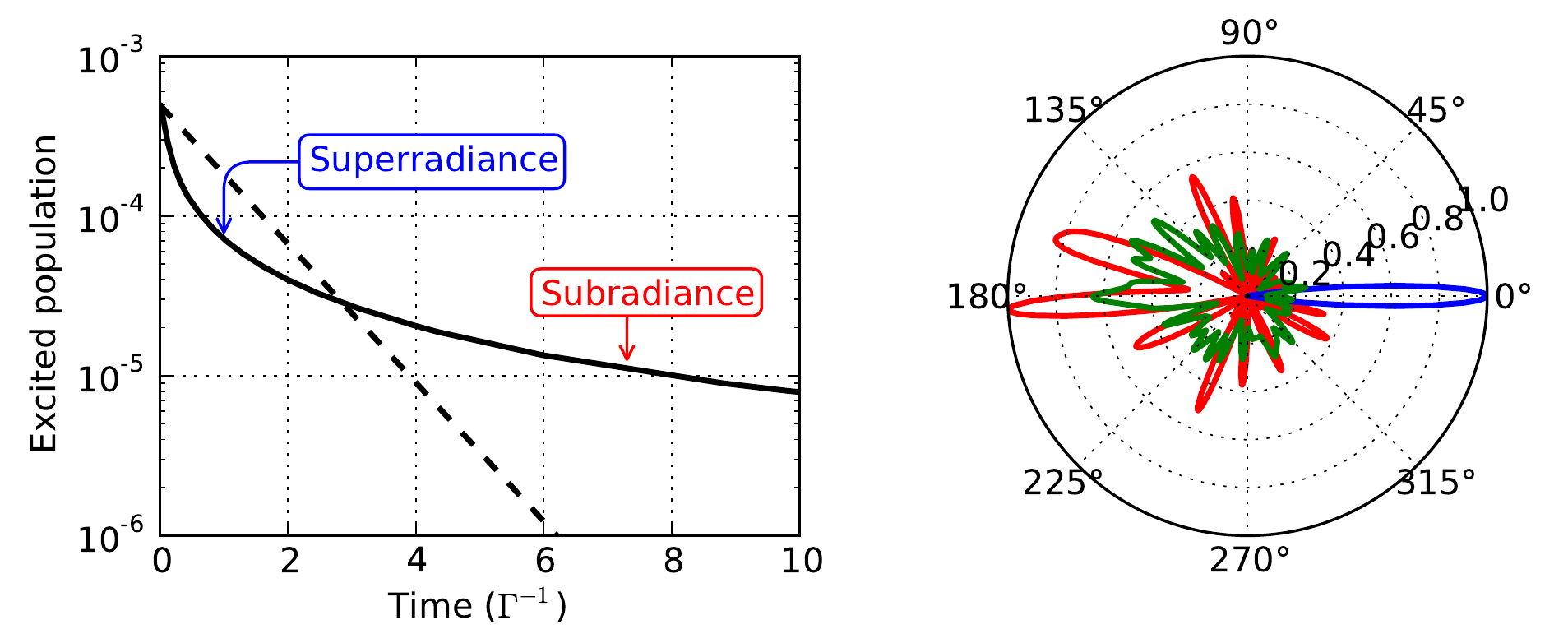}}}
\caption{(color online) Excited state population $P=\sum_j |\beta_j|^2$ decay after switching off the laser (the initial state corresponds
to the steady state. The black dashed curve shows the single atom decay (without cooperative effects). At first, the fast decay corresponds
to superradiance with a rate $\Gamma_N = (1 + b_0/12) \Gamma$. After some time, part of the light remains trapped in the cloud which
corresponds to subradiant emission. Parameters for the simulation: $N = 2000$, $k_0 \sigma_R = 10$, $\Omega_0 = 0.01 \, \Gamma$,
$\Delta_0 = 10 \, \Gamma$ (the laser was on before during $50 \, \Gamma^{-1}$ to let the system reach the steady state).
The superradiant emission diagram (blue curve) is strongly forward directed. The emission diagram of the subradiant states (red curve)
is isotropic. The green curve shows the subradiant emission diagram averaged over height realizations of disorder.}
\label{fig6}
\end{figure}

\begin{figure}[t]
\centerline{{\includegraphics[height=5cm]{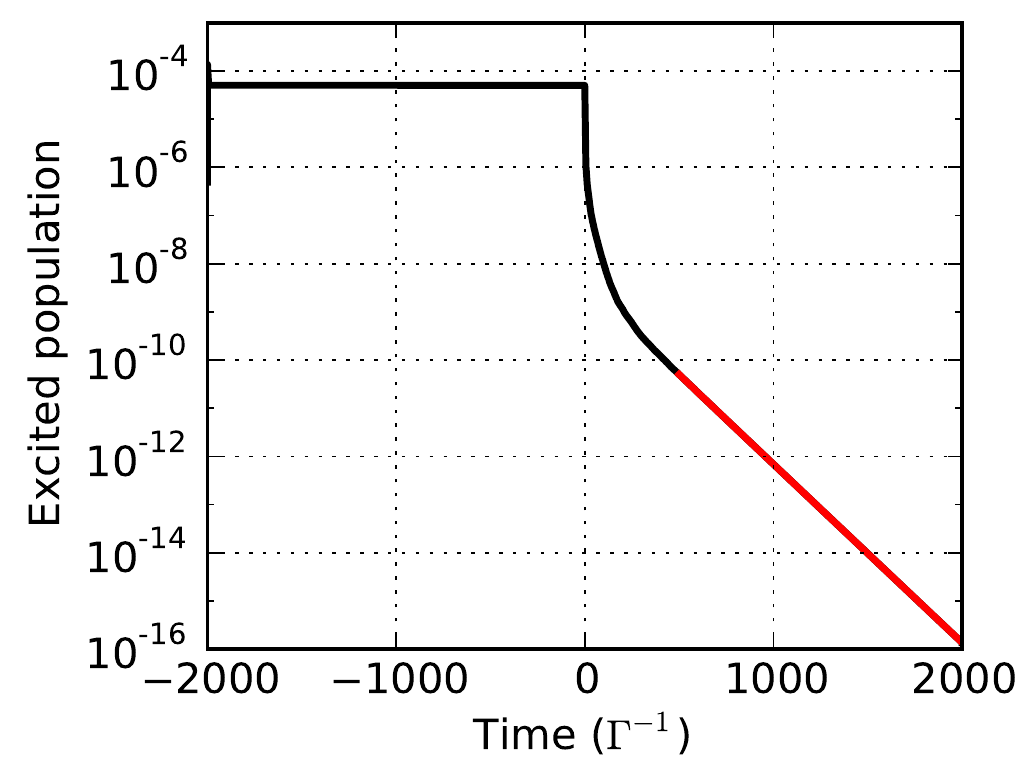}}}
\caption{(color online) Excited state population as a function of time for the same parameters as Fig. \ref{fig6} ($N = 2000$,
$k_0 \sigma_R = 10$, $\Omega_0 = 0.01 \, \Gamma$, $\Delta_0 = 10 \, \Gamma$). The laser is switched off at $t=0$.
The curve together with Fig. \ref{fig6} show that the subradiant decay is purely exponential only after a certain
amount of time (red part of the curve), when the mode with the longest lifetime dominates. We call this final decay
rate the subradiant decay rate. In this example, $\Gamma_{\text{sub}} = 8.5 \, 10^{-3} \, \Gamma$.}
\label{fig7}
\end{figure}

After driving the system with the laser for a long time, which allows populating subradiant states using the scheme sketched in Fig. \ref{fig5},
we monitor the decay of the system by looking at the excited state population $P(t)$ after switching off the laser. A typical decay curve of
the excited state population computed from the numerical solution of the effective Hamiltonian approach is shown on Fig. \ref{fig6}.
This figure shows the excitation probability $P(t)=\sum_j|\beta_j|^2$ as a function of time (black solid line), obtained by integrating
Eq. (\ref{beta}) for $N=2000$ atoms distributed by a Gaussian distribution with $k_0\sigma_R=10$. The other parameters are $\Omega_0=0.01\Gamma$
and $\Delta_0=10 \, \Gamma$ and the laser is switched off after $t=50 \, \Gamma^{-1}$. The origin of time is set such that it corresponds to the
time when the laser is switched off. Under the action of the continuous laser excitation, the atoms reach a quasi-stationary state close to the
timed Dicke state. The small subradiant fraction present in the atomic state after the exposition to the laser can be detected observing the
excitation decay after the laser has been switched off. The fast initial decay rate of the superradiant state is $\Gamma_N = (1 + b_0/12) \Gamma$
as expected since the steady state corresponds approximately to the timed Dicke state. After some time, the emission rate becomes much below the
single atom emission rate (black
dotted line in Fig. \ref{fig6}). It corresponds to the subradiant emission region. At first, the subradiant decay is not exponential since
several modes decay simultaneously. For longer times, it then ends up with a pure exponential decay, referred as the subradiant decay rate,
when only one long-lived mode dominates \cite{BienaimePhD,Bienaime12}, as shown in the red part of the decay curve of Fig. \ref{fig7}.
We have checked numerically the very intuitive result that the subradiant decay rate measured on the relaxation curves $P(t)$ corresponds
to the longest lifetime of the effective Hamiltonian modes \cite{BienaimePhD}.  This confirms the role of cooperativity in long lived
excitations in the cloud as investigated in ref. \cite{Akkermanns08}.

Using Eq. (\ref{Isca}) from the previous section, we can study the emission diagram of the system.
Fig. \ref{fig6} shows the emission diagram as a function of time during the decay of the cloud.
At first, the emission diagram of the timed Dicke state is clearly forward directed, a phenomenon reminiscent of Mie scattering.
At longer times, subradiant modes show isotropic emission diagrams: they do not possess the symmetry of the laser excitation since
they are not directly coupled to it. This property can be exploited in the experimental detection of subradiance.

In Ref.\cite{Bienaime12}, we proposed to use inhomogeneous broadening schemes such as the cloud optical thickness, the cloud
temperature, or the driving laser intensity as possible control parameters for subradiance. However, other parameters such as a far
detuned speckle field, magnetic fields, or near field couplings can also be used. By control of subradiance, we mean two different things:
controlling the population of the subradiant states as well as their decay rates. Exploiting these inhomogeneous broadening schemes allow
to control and tune the dipole-dipole couplings, which is the genuine interaction leading to cooperative effects (superradiance, subradiance,
cooperative Lamb shift). This would allow for the first observation of subradiant emission from a cloud of atoms in free space.

\subsection{Dipole-dipole induced suppression of excitation}

In the perturbative limit, where each dipole is driven by
the external field, plus a small perturbation by the field
scattered by all other dipoles, we can obtain an analytical solution
of the excitation state population and the angle-resolved
scattered field.

\begin{figure}[t]
\centering{\includegraphics[width=\linewidth, height=5cm]{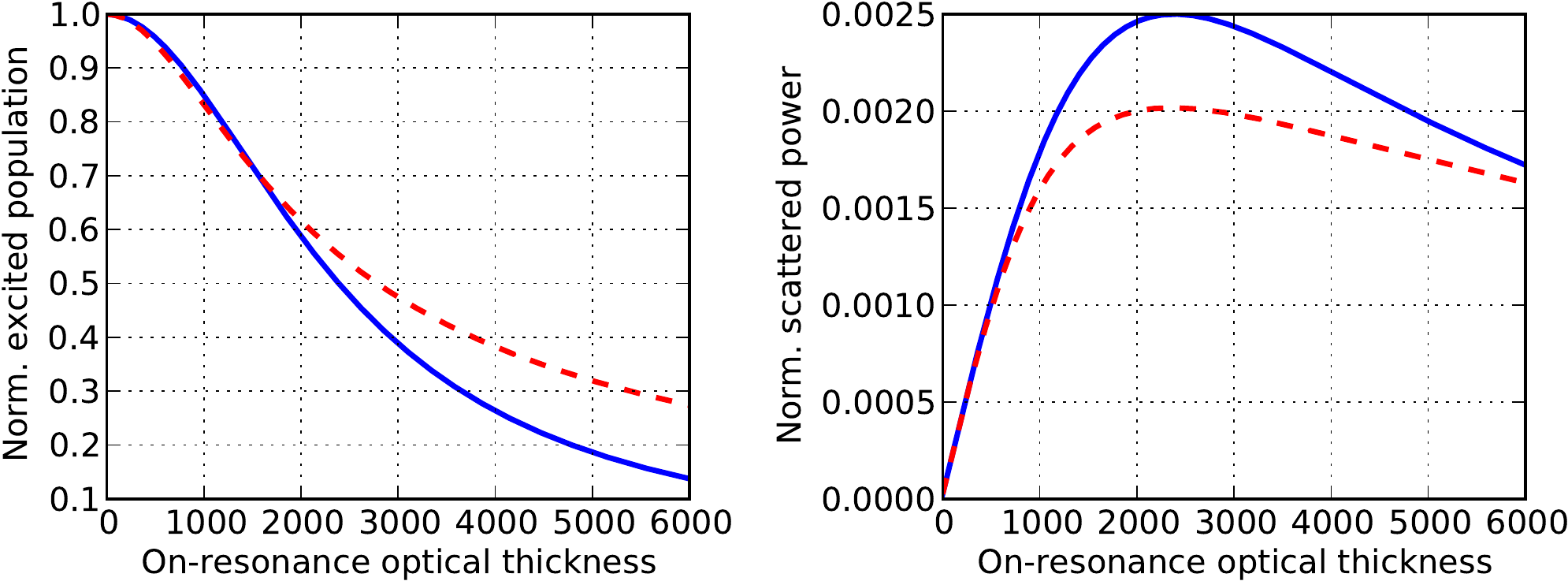}}
\caption{(color online) Normalized excited state population (left) and normalized
total scattered power (right) as a function of the optical thickness $b_0$ for $\Delta_0=100 \, \Gamma$.
The blue solid lines correspond to Eqs. (\ref{Pnorm}), (\ref{Psnorm}) and the red dashed lines to Eqs. (\ref{PRA1}), (\ref{PRA2}).} \label{fig8}
\end{figure}

>From Eq. (\ref{Isca}) we obtain the steady-state scattered
intensity in the direction $(\theta,\phi)$ for the timed Dicke
state, still neglecting saturation:
\begin{equation}\label{IscaTD}
    I(r,\theta,\phi)=\left(\frac{I_0}{16\pi^2 k_0^2 r^2}\right)\frac{\Gamma^2 N^2|S_N(\theta,\phi)|^2}{4\Delta_N^2+\Gamma_N^2}.
\end{equation}
and the total scattered intensity
\begin{equation}\label{PscaTD}
    P_{s}=r^2 \int_0^{2\pi}d\phi\int_0^\pi d\theta\sin\theta \, I(r,\theta,\phi)
    =P_0\frac{N\Gamma\Gamma_N}{4\Delta_N^2+\Gamma_N^2},
\end{equation}
 where $I_0$ is the incident
intensity and $P_0=I_0/(4\pi k_0^2)$. In Fig. \ref{fig8}, we plot the normalized excited state population (blue solid line on the left figure) as a function of $b_0$ and for $\Delta_0=100\Gamma$,
\begin{equation}\label{Pnorm}
    \frac{P}{NP^{(1)}}=\frac{4\Delta_0^2+\Gamma^2}{4\Delta_N^2+\Gamma_N^2}\approx
    \frac{4\Delta_0^2+\Gamma^2}{4\Delta_0^2+\Gamma^2(1+b_0/12)^2},
\end{equation}
and the normalized total scattered power (blue solid line on the left figure)
\begin{equation}\label{Psnorm}
    \frac{P_s}{NP_0}\approx
    \frac{\Gamma^2}{4\Delta_0^2+\Gamma^2(1+b_0/12)^2}(1+b_0/12),
\end{equation}
where $P^{(1)}=\Omega_0^2/(4\Delta_0^2+\Gamma^2)$ is the
single-atom excitation probability and we have neglected the
collective Lamb shift, $\Delta_N\approx \Delta_0$. We observe that
increasing the optical thickness $b_0$ the excitation population
decreases, whereas the total scattered power has a maximum around
$b_0\sim 24(\Delta_0/\Gamma)$. The normalized excited state population and total scattered power resulting from the partial wave solution (\ref{betaBTD}) in the continuous-density
approximation, obtained in ref.\cite{Bachelard11} for large Gaussian clouds,
\begin{equation}\label{PRA1}
    \frac{P}{NP^{(1)}}=
    \frac{4\Delta_0^2+\Gamma^2}{\Delta_0\Gamma(b_0/3)}\arctan\left[\frac{\Delta_0\Gamma(b_0/3)}{4\Delta_0^2+\Gamma^2(1+b_0/6)}\right],
\end{equation}
and
\begin{equation}\label{PRA2}
    \frac{P_s}{NP_0}=
    \frac{3}{b_0}\ln\left[1+\frac{\Gamma^2
    b_0}{3}\frac{1+b_0/12}{4\Delta_0^2+\Gamma^2}\right],
\end{equation}
are also plotted in Fig.\ref{fig8} (red dashed lines): they show a
deviation from the timed Dicke approximation for very large
resonant optical thickness $b_0$.

>From these curves we can see that even for dilute clouds of cold
atoms, the long range coupling between the dipoles leads to a
cooperative modification of the excitation of the atoms and its
related total scattered power. Note that the total scattered power initially
increases with increasing number of atoms, despite the decrease in
the normalized total population of the excited state. This can be
understood by the fact that cooperativity leads to enhanced
superradiant emission rates. For larger number of atoms, the
suppression of the atomic excitation dominates the enhanced
superradiant emission and the total scattering rate of the large
cloud of atoms is reduced by the dipole-dipole couplings.  We
stress that even though the signature of such a suppression of
fluorescence of the cloud of $N$ atoms might bear a resemblance
with a photon blockade regime \cite{Tong, Singer, Molmer}, our
model does not take into account optical nonlinearities required
to describe such photon-photon coupling. A suppression of
excitation can thus be obtained in the absence of nonlinear
optical response. In contrast to dipole blockade effects in
Rydberg states with near-field ($1/r^3$) or Van der Waals
coupling ($1/r^6$), we are in the presence of long-range
dipole-dipole couplings where all atoms participate in the
suppression of the atomic excitation, and not only a small volume
around an excited atom.

\section{Conclusion}

In this paper, we have presented a master equation and an effective Hamiltonian approach to describe
cooperative effects in clouds of cold atoms. This master equation approach, even though still restricted
to single excitation, allows to go beyond the effective Hamiltonian approach. In particular, we have highlighted the possibility of cooperative suppression of the atomic excitation, via the long range dipole-dipole couplings and in the absence of any non-linear photon blockade mechanism. Future work will include the possibility of experimental observation of Dicke subradiance, long range dipole-dipole blockade and cooperative effects beyond linear optics.

\begin{acknowledgement}
  Funding from IRSES project COSCALI and from USP/COFECUB is acknowledged.
\end{acknowledgement}


\begin{thebibliography}{[10]}

\bibitem{Dicke54}R.~H. Dicke, Phys. Rev. \textbf{93}, 99 (1954).
\bibitem{Javanainen} J. Javanainen, Phys. Rev. Lett. \textbf{72}, 2375 (1994).
\bibitem{Raimond} J. M. Raimond, P. Goy, M. Gross, C. Fabre, S. Haroche, Phys. Rev. Lett. \textbf{49}, 117 (1982).
\bibitem{Trammell99} J. P. Hannon, G. T. Trammell, Hyperfine Interactions \textbf{123/124}, 127 (1999).
\bibitem{Rohlsberger10} R. R\"ohlsberger, K. Schlage, B. Sahoo, S. Couet, R. R\"uffer, Science \textbf{328}, 1248-1251 (2010).
\bibitem{Scully06} M. O. Scully, E. S. Fry, C. H. R. Ooi, K. Wo\'dkiewicz, Phys. Rev. Lett. \textbf{96}, 010501 (2006).
\bibitem{Eberly06}J. H. Eberly, J. Phys. B: At. Mol. Opt. Phys. \textbf{39}, S599 (2006).
\bibitem{Svidzinsky08} A. A. Svidzinsky, J.-T. Chang,  M.~O. Scully, Phys. Rev. Lett. \textbf{100}, 160504 (2008).
\bibitem{Svi10} A. A. Svidzinsky, J.-T. Chang, M.~O. Scully, Phys. Rev.
A \textbf{81}, 053821 (2010).
\bibitem{Courteille10} Ph. W. Courteille, S. Bux, E. Lucioni, K. Lauber, T.
Bienaim\'e, R. Kaiser, N. Piovella, Eur. Phys. J. D \textbf{58},
69 (2010).
\bibitem{Bienaime10} T. Bienaim\'e, S. Bux, E. Lucioni, Ph. W. Courteille, N. Piovella, R. Kaiser, Phys. Rev. Lett. \textbf{104}, 183602 (2010).
\bibitem{Bender10} H. Bender, C. Stehle, S. Slama, R. Kaiser, N. Piovella, C. Zimmermann, Ph. W.
Courteille, Phys. Rev. A \textbf{82}, 011404 (2010).
\bibitem{Bienaime11} T. Bienaim\'e, M. Petruzzo, D. Bigerni, N. Piovella, R. Kaiser, J. Mod. Opt. \textbf{58}, 1942 (2011).
\bibitem{Popescu} M. Tiersch, S. Popescu, H. J. Briegel1, arXiv:1104.3883v.
\bibitem{Bariani} F. Bariani, T. A. B. Kennedy, Phys. Rev. A \textbf{85}, 033811 (2012).
\bibitem{Tong} D. Tong, S. M. Farooqi, J. Stanojevic, S. Krishnan, Y. P. Zhang, R. C\^ot\'e, E. E. Eyler, P. L. Gould, Phys. Rev. Lett.
\textbf{93}, 063001 (2004).
\bibitem{Singer} K. Singer, M. Reetz-Lamour, T. Amthor, L. G. Marcassa, M. Weidem\"uller, Phys. Rev. Lett. \textbf{93}, 163001 (2004).
\bibitem{Molmer} M. Saffman, T. G. Walker, K. Molmer, Rev. Mod. Phys. \textbf{82}, 2313 (2010).
\bibitem{Agarwal} G. S. Agarwal, \textit{Quantum Statistical Theories of Spontaneous Emission and their Relation to other
Approaches}, Springer tract in Modern Physics, ed. G. H\"{o}hler,
Springer-Verlag, Berlin (1974).
\bibitem{Das} S. Das, G. S. Agarwal, M. O. Scully, Phys. Rev. Lett. \textbf{101}, 153601 (2008).
\bibitem{Scully09LS} M.~O. Scully, Phys. Rev. Lett. \textbf{102}, 143601 (2009).
\bibitem{Friedberg} R. Friedberg, S. R. Hartman, J. T. Manassah, Phys. Rep. \textbf{7}, 101 (1973).
\bibitem{Fano} U. Fano, Phys. Rev. \textbf{124}, 1866 (1961).
\bibitem{Cummings} F. W. Cummings, Phys. Rev. A \textbf{33}, 1683 (1986).
\bibitem{Scully09}  M.~O. Scully and A.~A. Svidzinsky, Phys. Rev. Lett.  \textbf{373}, 1283 (2009).
\bibitem{Bienaime12} T. Bienaim\'e, N. Piovella, R. Kaiser, Phys. Rev. Lett. \textbf{108}, 123602 (2012).
\bibitem{Gaetan} A. Ga\"etan, Y. Miroshnychenko, T. Wilk, A. Chotia, M. Viteau, D.
Comparat, P. Pillet, A. Browaeys, P. Grangier, Nature Phys. \textbf{5}, 115 (2009).
\bibitem{BienaimePhD} T. Bienaim\'e, Ph.D. thesis, Universit\'e de Nice Sophia Antipolis (2011).
\bibitem{Bachelard11} R. Bachelard, N. Piovella, Ph. W. Courteille, Phys. Rev. A  \textbf{84}, 013821 (2011).
\bibitem{Manassah12} J. T. Manassah, Phys. Rev. A \textbf{85}, 015801 (2012).
\bibitem{Bachelard12} R. Bachelard,  Ph.  W. Courteille, R. Kaiser, N. Piovella, EPL \textbf{97}, 14004 (2012).
\bibitem{Brewer96} R. G. DeVoe and R. G. Brewer, Phys. Rev. Lett. \textbf{76}, 2049 (1996).
\bibitem{Pillet85} D. Pavolini, A. Crubellier, P. Pillet, L. Cabaret, S. Liberman, Phys. Rev. Lett. \textbf{54}, 1917 (1985).
\bibitem{Akkermanns08} E. Akkermanns, A. Gero, R. Kaiser, Phys. Rev. Lett. \textbf{101},  103602 (2008).

\end{thebibliography}
\end{document}